%% file: main_quantum_method.tex
\PassOptionsToPackage{super,sort&compress}{natbib}
\documentclass[aps,pra,onecolumn,11pt,tightenlines,raggedbottom,superscriptaddress,notitlepage,nofootinbib]{revtex4-2}

\usepackage{amsmath,amssymb}
\usepackage{booktabs,tabularx,array}
\usepackage{graphicx}
\usepackage{xcolor}
\usepackage{microtype}
\usepackage{placeins}
\usepackage{setspace}
\usepackage{etoolbox}
\makeatletter
\let\switch@array\relax
\makeatother
\setcitestyle{super,comma}
\definecolor{referenceblue}{RGB}{0,70,170}
\usepackage[colorlinks=true,linkcolor=referenceblue,citecolor=referenceblue,urlcolor=referenceblue]{hyperref}
\AtBeginDocument{}
\apptocmd{\appendix}{\counterwithout{equation}{section}}{}{}
\makeatletter
\AtBeginDocument{\renewcommand{\p@subsection}{}}
\makeatother

\begin{document}

\title{Agent-assisted experimental strategy transfer across quantum processors}
\author{Shi-Yao Hou}
\affiliation{College of Physics and Electronic Engineering, Center for Computational Sciences, Sichuan Normal University, Chengdu 610068, China}
\author{Zipeng Wu}
\affiliation{Shenzhen SpinQ Technology Co., Ltd., Shenzhen 518043, China}
\author{Chao Zhang}
\affiliation{Shenzhen SpinQ Technology Co., Ltd., Shenzhen 518043, China}
\author{Guanru Feng}
\affiliation{Shenzhen SpinQ Technology Co., Ltd., Shenzhen 518043, China}
\author{Keren Li}
\email{likr@szu.edu.cn}
\affiliation{College of Physics and Optoelectronic Engineering, Shenzhen University, Shenzhen 518060, China}
\author{Jingen Xiang}
\affiliation{Shenzhen SpinQ Technology Co., Ltd., Shenzhen 518043, China}

\begin{abstract}
Progress in quantum hardware requires a growing volume of experiments across diverse devices, straining the capacity of human experts. Automation remains challenging because device differences and drift can render workflows developed on one processor unsuitable for another. Here we present an agent-assisted experimental strategy for conducting experiments across three-qubit nuclear magnetic resonance processors. Developed on one processor and applied sequentially to four others, the strategy uses an agent to assess experimental evidence, select subsequent measurements and adapt procedures, with operators providing instrument preparation and physical intervention.
Using three-qubit Grover state preparation as a common task, each deployment combined device-specific calibration, control-pulse optimization and measurement. 
The experimental strategy was refined through sequential deployment, and all four processors ultimately produced outputs dominated by their respective marked states. The marked-state populations of the reconstructed dominant eigenstates ranged from \(0.910\) to \(0.959\), with relative deviations below \(3.8\%\) from the ideal value of 0.945. These results demonstrate that an agent-assisted experimental strategy can be reused and refined across processors to realize a common quantum task, providing a basis for scaling experimental practice beyond individual devices.
\end{abstract}
\maketitle

\section*{Introduction}
The development and operation of quantum hardware increasingly require experiments to be coordinated or repeated across devices. This need is especially apparent in distributed quantum computing and quantum networking \citep{pompili2022network,main2025distributed}, parallel execution across multiple quantum processors \citep{li2025thread}, and benchmarking processors, modules and network links \citep{proctor2022capabilities,zhu2022crossplatform,helsen2023network}. Transferring an experimental workflow between processors is nevertheless difficult: even nominally similar devices differ, and their behaviour may drift over time~\citep{proctor2020drift}. Experimental workflows often rely on device-specific assumptions and settings, making local adaptation necessary when moving between processors.

Expert experimentalists provide this adaptability by adjusting procedures to the device and judging whether the available evidence supports proceeding or calls for further investigation. To reduce manual intervention, machine-learning methods have automated aspects of device tuning and real-time feedback control \citep{moon2020tuning,reuer2023feedback}, while autonomous laboratories can integrate experiment selection, execution and analysis in closed-loop workflows \citep{szymanski2023autonomous,canty2025sdl}. Yet automation tied to a fixed workflow may fail as hardware changes. The challenge is therefore to formalize an adaptable experimental strategy that can be applied across devices.

Recent advances in artificial intelligence (AI) agents are extending autonomous laboratories towards workflows that proceed from a high-level research objective through experimental planning and execution to analysis \citep{boiko2023autonomous}. Within quantum computing, AI is increasingly being applied across the stack, including in quantum-device design and operation \citep{alexeev2025aiquantum}. At the experimental level, recent studies have deployed language-model agents for superconducting-processor operation and calibration \citep{cao2025agents,li2026llmqubit,xu2026vibe}, safety-gated trapped-ion control \citep{wang2026artiq} and quantum sensing \citep{isogawa2026agentic}. Some have also explored calibration and control-interface portability across experimental platforms \citep{xu2026vibe,wang2026artiq}. Moving from calibration and control portability to task realization, we ask: \emph{Can an agent-assisted strategy implement a common multiqubit task across processors by adapting experimental procedures to local device conditions and evidence?}

Nuclear magnetic resonance (NMR) provides an established setting for quantum control and computation~\citep{vandersypen2005nmr}, including experimental polynomial optimization on a four-qubit processor~\citep{li2021polynomial}. Here we evaluate an agent-assisted experimental strategy on five three-qubit NMR processors, using three-qubit Grover target-state preparation as a common experimental task \citep{grover1997,das2003grover}. The processors shared a common experimental interface but differed in Hamiltonian parameters, spectral structure and operating conditions. The strategy was developed under human supervision on one processor and applied sequentially to four others. Each deployment involved local calibration and measurement, together with device-specific control-pulse optimization using gradient ascent pulse engineering (GRAPE) \citep{khaneja2005grape}. Across the four deployment processors, the marked-state populations of the reconstructed dominant eigenstates ranged from \(0.910\) to \(0.959\), with relative deviations below \(3.8\%\) from the ideal value of \(0.945\).
Experience gained during sequential deployment informed subsequent strategy refinements. Together, these results demonstrate the feasibility of reusing and adapting an agent-assisted experimental strategy across a quantum processor family to realize a common multiqubit task.

\section*{Results}

\subsection*{Agent-guided framework for strategy transfer}
\begin{figure}[t]
\centering
\includegraphics[width=\linewidth]{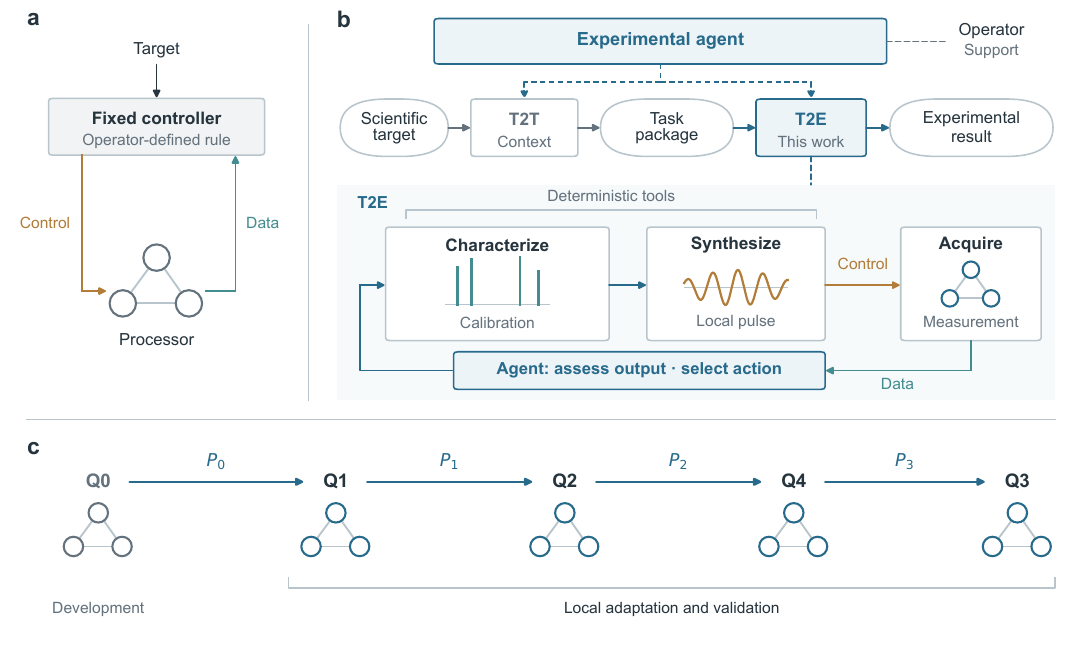}
\caption{\textbf{Agent coordination and strategy transfer.} \textbf{a, Fixed-rule feedback.} A controller updates predefined control variables. \textbf{b, Agent-guided realization.} The framework connects a scientific target to an experimental result through target-to-task (T2T; context only) and task-to-experiment (T2E; evaluated here). In T2E, a single agent invokes deterministic tools for characterization, control synthesis and acquisition, assesses their outputs and selects subsequent actions, with operators providing instrument preparation and physical intervention. \textbf{c, Strategy transfer and local realization.} Q0 supports guided development; Q1, Q2, Q4 and Q3 undergo sequential local adaptation and validation. \(P_i\) labels the strategy entering the next processor; the Q3 entry version \(P_3\) was inferred from preserved records.}
\label{fig:workflow}
\end{figure}

We designed an agent-guided framework that separates a transferable experimental strategy from its device-specific physical realization (Fig.~\ref{fig:workflow}). In the fixed-rule feedback scheme shown in Fig.~\ref{fig:workflow}a, measurements update predefined control variables. 
The conceptual framework links a scientific target to an experimental result through two stages: target-to-task (T2T) and task-to-experiment (T2E) (Fig.~\ref{fig:workflow}b). T2T translates the scientific target into a hardware-independent task specification comprising a target state or observable, a logical circuit and an evaluation criterion. Recent language-model systems have addressed components of T2T, including quantum-code and circuit generation \citep{dupuis2024qiskit,jern2025agentq} and experimental-blueprint generation from classes of target states \citep{arlt2026metadesign}; T2T is not evaluated here. We focus on T2E, which realizes a supplied task specification on physical hardware.

In T2E, the agent invokes and coordinates deterministic tools to characterize the processor, synthesize controls, acquire measurements and reconstruct experimental outputs. It uses the resulting evidence to select subsequent actions and adapt the experimental procedure. Operators retain responsibility for instrument preparation, physical adjustment, safety and procedural authorization. The agent therefore coordinates decisions at a higher level than a conventional controller, while numerical calculations remain assigned to deterministic tools and physical interventions to operators. 

To evaluate strategy transfer across processors, we developed the strategy under human supervision on Q0 and deployed it sequentially to Q1, Q2, Q4 and Q3 (Fig.~\ref{fig:workflow}c). All deployments used a common three-qubit Grover task and experimental interface. Each processor required local calibration, measurement references and control-pulse optimization, with the agent guiding adaptation based on local experimental evidence. This design allowed us to examine how an experimental strategy developed on one processor could be applied and adapted to others.

\subsection*{Developing the initial strategy}

\begin{figure}[t]
\centering
\includegraphics[width=\linewidth]{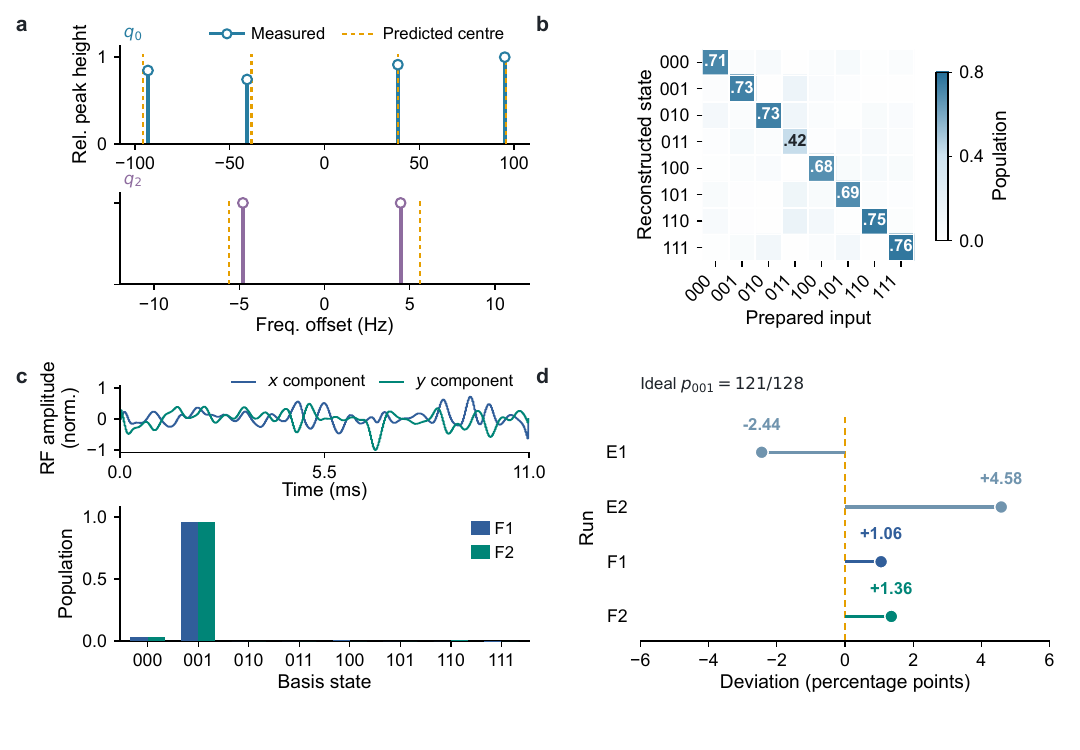}
\caption{\textbf{Development of the experimental strategy on Q0.} \textbf{a}, Four \(q_0\) peaks and the two inner \(q_2\) peaks, with measured relative heights; dashed lines mark Hamiltonian-predicted frequencies. Frequencies are relative to fitted resonance offsets \(\Delta\nu_i\); \(q_i\) labels qubits within Q0. \textbf{b}, Empirical response matrix combining preparation, measurement and reconstruction effects. \textbf{c}, Normalized \(x/y\) radiofrequency (RF) control components of the 11-ms pulse and final-run populations (F1/F2). \textbf{d}, Early (E1/E2) and final (F1/F2) deviations, \(100(p_{001}-121/128)\), in percentage points, calculated from unrounded populations. Horizontal segments connect each run to zero deviation (dashed line). All four runs used the same pulse and response matrix.}
\label{fig:q0}
\end{figure}

On the development processor Q0, an agent coordinated the three-qubit
Grover experiment under operator supervision to establish the initial
experimental strategy \(P_0\) (Fig.~\ref{fig:q0}).

We first estimated the local Hamiltonian parameters by fitting resonance lines from \(x\)- and \(y\)-phase excitation spectra, with four assigned lines per spin. The root-mean-square and maximum absolute fitting errors were 1.186 and 2.390~Hz, respectively; representative measured and fitted line positions are shown in Fig.~\ref{fig:q0}a. We then prepared the eight computational-basis inputs and reconstructed their population distributions to form an \(8\times8\) empirical response matrix. This matrix captured the combined effects of preparation, measurement and reconstruction, with the intended input state having the largest population in every column (Fig.~\ref{fig:q0}b).

An 11-ms radio-frequency (RF) pulse optimized using GRAPE implemented the state-preparation task defined by two Grover iterations, whose ideal marked-state population is \(121/128\simeq0.945\). Fig.~\ref{fig:q0}c
shows the pulse and the final population distributions obtained by tomography and response-matrix correction. Two early runs (E1/E2) and two final runs (F1/F2), all using the same pulse and response matrix, yielded marked-state populations of 0.921, 0.991, 0.956 and 0.959, respectively (Fig.~\ref{fig:q0}d). E1/E2 were interleaved with an extended basis-state calibration block, whereas F1/F2 were acquired in a shorter, closely paired block. The final pair showed closer agreement, with a population difference of 0.003.

This operator-guided development provided the basis for \(P_0\), which defined rules for action selection and stopping. These rules specified that unreliable spectral fits required further characterization, while insufficient reference data or inconsistent repeated measurements required additional acquisition before task performance could be assessed. The strategy assigned routine experimental decisions to the agent and called for operator assistance when physical conditions needed stabilization.

\subsection*{Strategy transfer across four processors}

Following development on Q0, we deployed the T2E strategy sequentially to Q1, Q2, Q4 and Q3 (Fig.~\ref{fig:workflow}c). At each processor, the agent assessed local device information and calibration results, then invoked the tools needed for characterization and control-pulse optimization. Each deployment implemented the state-preparation task defined by two three-qubit Grover iterations, with a processor-specific marked computational-basis state.

Transfer required adapting the procedure to the evidence available on each processor. On Q1, the agent rejected a fit that relied on settings from the development processor and requested fresh local measurements. On Q2, unstable reference signals prompted additional checks and updated references before proceeding. On Q4, overlapping spectral signals made the original analysis procedure unsuitable, so the agent revised the method and subsequently reused it on Q3 with new local data. Across these deployments, the agent used experimental evidence to decide whether to collect more data, revise the procedure, request operator assistance or stop. These observations and responses are summarized in Supplementary Table~\ref{tab:supp_agent_cases}, with the Q4/Q3 method adaptation detailed in Supplementary Sec.~\ref{supp:adaptation_case} and Supplementary Fig.~\ref{fig:supp_adaptation}.

\begin{figure}[t]
\centering
\includegraphics[width=\linewidth]{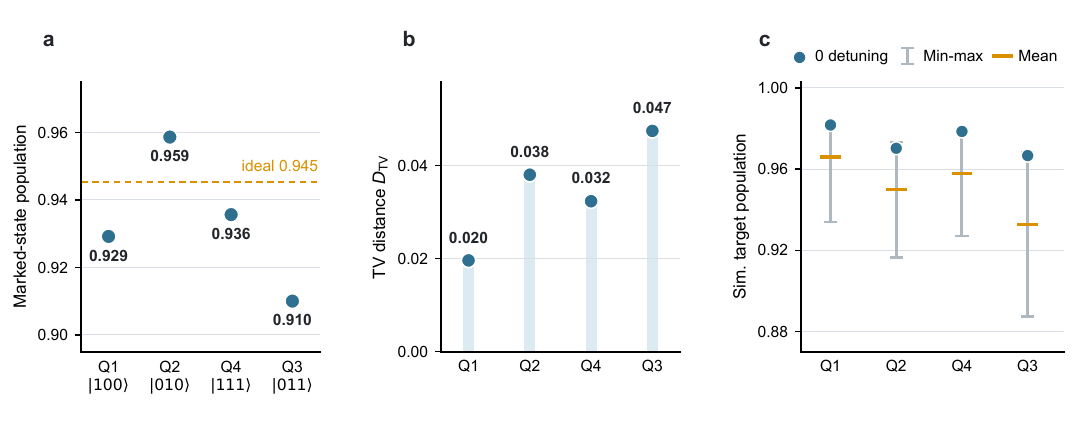}
\caption{\textbf{Grover outcomes after cross-processor transfer.} \textbf{a,b}, Marked-state populations and total-variation (TV) distances from ideal population distributions. The dashed line in \textbf{a} marks the ideal population, \(121/128\simeq0.945\). \textbf{c}, Simulated marked-state populations over the pulse-optimization ensemble, comprising five detunings from \(-40\) to \(40\)~Hz in 20-Hz steps and three radiofrequency-amplitude scales (0.95, 1.00 and 1.05). Vertical bars span the minimum to maximum, horizontal ticks mark the mean, and circles indicate zero detuning at unit amplitude scale.}

\label{fig:cross}
\end{figure}

All four processors produced outputs dominated by their respective marked states. The marked-state populations of the reconstructed dominant eigenstates were 0.929, 0.959, 0.936 and 0.910 for Q1, Q2, Q4 and Q3, respectively, with relative deviations below \(3.8\%\) from the ideal value of \(121/128\simeq0.945\) (Fig.~\ref{fig:cross}a). The corresponding total-variation (TV) distances from the ideal eight-state population distributions ranged from 0.020 to 0.047 (Fig.~\ref{fig:cross}b). Across the detuning and RF-amplitude ensemble used for pulse optimization, the minimum simulated marked-state populations ranged from 0.887 to 0.934 across processors (Fig.~\ref{fig:cross}c).

Together, the recorded method adaptations and Grover outcomes support the reuse and refinement of an agent-assisted experimental strategy across this NMR processor family.

\subsection*{Agent-guided repeated measurements}

After the initial transfer, we assessed the repeatability of Grover outputs in subsequent experimental sessions. Five completed sessions
across Q2, Q4 and Q3 each included reference measurements and three complete tomography repeats, yielding 15 complete measurements in total.

\begin{figure}[t]
\centering
\includegraphics[width=\linewidth]{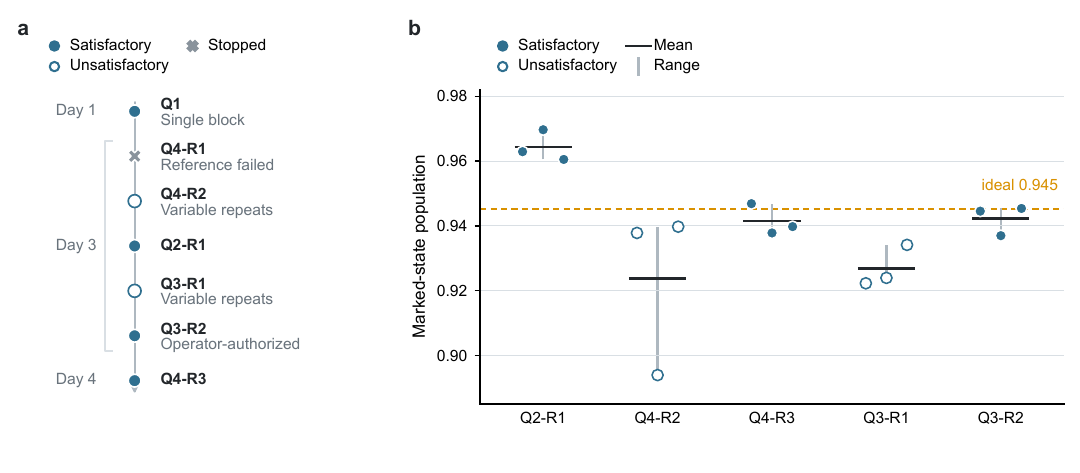}
\caption{\textbf{Agent-guided repetition and output stability.} \textbf{a}, Session acquisition order; days are relative to the first session. Q1 shows the earlier single-block validation; Q4-R1 stopped after reference failure. \textbf{b}, All 15 complete seven-setting repeats, ordered left to right within each session. For sessions with three repeats, filled and open circles in both panels indicate satisfactory and unsatisfactory repeatability, respectively, assessed qualitatively from their variation. Horizontal lines and vertical bars show each session's mean and range.}
\label{fig:replication}
\end{figure}

The agent used reference checks to guide whether acquisition could proceed. Q4-R1 stopped before target acquisition because its reference spectra failed the device-readiness checks. Following recovery from a server interruption, the agent coordinated completion of Q3-R1 and proposed an additional session, Q3-R2, which was authorized by the operator (Fig.~\ref{fig:replication}a).

Repeatability varied across sessions. The marked-state populations of the reconstructed dominant eigenstates spanned 0.894--0.940 in Q4-R2 and 0.938--0.947 in the later Q4-R3 session. On Q3, the corresponding
ranges were 0.922--0.934 in Q3-R1 and 0.937--0.945 in Q3-R2. Q2-R1 yielded populations between 0.960 and 0.970. In a retrospective qualitative assessment of within-session variation, Q4-R2 and Q3-R1 were judged unsatisfactory, whereas Q2-R1, Q4-R3 and Q3-R2 were judged satisfactory. Fig.~\ref{fig:replication}b shows all 15 repeats, together with each session's mean and observed range.

These sessions illustrate how reference checks, interruption recovery and repeated measurements supported continued experimentation while revealing variation in task outputs.

\section*{Discussion}
Transferring a quantum experiment requires connecting a common logical task to a new physical realization. Here, a strategy developed under human supervision on one three-qubit NMR processor was reused and refined on four others, each with a locally rebuilt Grover experiment. The recorded actions show how the agent coordinated this process: uncertain fits or inconsistent references were followed by additional measurements, and the fitting procedure was adapted to the observed spectral structure.

Recent studies demonstrate gate calibration and Greenberger--\allowbreak Horne--\allowbreak Zeilinger state preparation \citep{cao2025agents}, on-demand experimental code generation \citep{li2026llmqubit}, and autonomous sensing with evaluations of scientific judgement \citep{isogawa2026agentic}. They focus on constructing, executing and assessing experiments, but do not examine strategy transfer for a common multiqubit task across processors requiring local adaptation of physical models, measurement references and controls.

More closely related work addresses calibration and control across devices. Vibe Calibration distils expert practice into reusable skills and demonstrates autonomous single-qubit characterization at scale. Its separate chip-transfer test examines whether fine-tuned agents follow a newly supplied calibration workflow \citep{xu2026vibe}. Wang et al.~\citep{wang2026artiq} demonstrate calibration and magnetic-field compensation in a trapped-ion experiment, with agent-generated control code checked and authorized before execution. Their deployment on a second apparatus tests control-interface compatibility on different hardware, without active ion feedback. These studies focus on reusing calibration workflows and enabling operation on additional devices, but do not test whether a common multiqubit task can be realized after rebuilding each processor's physical model, measurement references and controls.

Our study extends cross-device calibration and control research to the realization of a common multiqubit task across distinct processors. With the logical task and experimental interface held common, each deployment rebuilt its Hamiltonian model, measurement references and Grover pulse. This setting reveals how the experimental procedure changes when new hardware alters both numerical parameters and the applicability of characterization and analysis methods. The records document how the strategy was adapted across processors, while the validation experiments characterize the task performance achieved by each local implementation.

Meta-design generates programs for families of quantum experiments \citep{arlt2026metadesign}, corresponding to design generation in the T2T stage shown in Fig.~\ref{fig:workflow}b. Our study focuses on the subsequent T2E stage, realizing a given task across devices. Integrating these stages could connect scientific targets to physical experiments through automated design generation and adaptation to local hardware. Extending the experimental strategy to superconducting or trapped-ion systems would involve adapting both the strategy and its supporting tools to each platform's physical models, measurement procedures and control capabilities. Testing and refining these strategies on additional platforms could allow experience gained on individual devices to guide a wider range of experimental tasks.
\section*{Methods}

\subsection*{Experimental platform}
Experiments were performed on five three-qubit liquid-state NMR processors (TriangulumPro, SpinQ Technology Co., Ltd.). The three inequivalent \(^{19}\mathrm{F}\) nuclei in iodotrifluoroethylene (\(\mathrm{C_2F_3I}\)) encoded the computational qubits, while a separate \(^{1}\mathrm{H}\) reference channel supported magnetic-field tracking. The instruments lacked active temperature control of the sample and magnet.

Experiments were initialized in an effective \(|000\rangle\) pseudo-pure state (PPS), prepared by alternating basis permutations with longitudinal relaxation \citep{feng2022triangulum}. Radiofrequency control and Fourier-transformed free-induction signals were used for spin manipulation and readout. Same-run PPS reference spectra provided per-spin amplitude and phase normalization. Processor identifiers and supporting platform characterization are provided in Supplementary Sec.~\ref{supp:platform}.

\subsection*{Agent implementation and transfer constraints}

The agent, identified in the Q0 manifest as Codex based on GPT-5, used Python experimental tools and a locally bundled \texttt{spinqlink} instrument interface. Its policy linked observations and quality checks to action selection, tool outputs and human interventions. Transfer context admitted generic policies and tools but excluded development-device numerical parameters, waveforms and reference shapes. These were procedural restrictions; automatic context isolation was not established. Configuration records, decision examples and the incomplete Q3 entry-context evidence are described in Supplementary Sec.~\ref{supp:agent}.

\subsection*{Grover task and ideal output}

\begin{figure}[tbp]
\centering
\includegraphics[width=\linewidth]{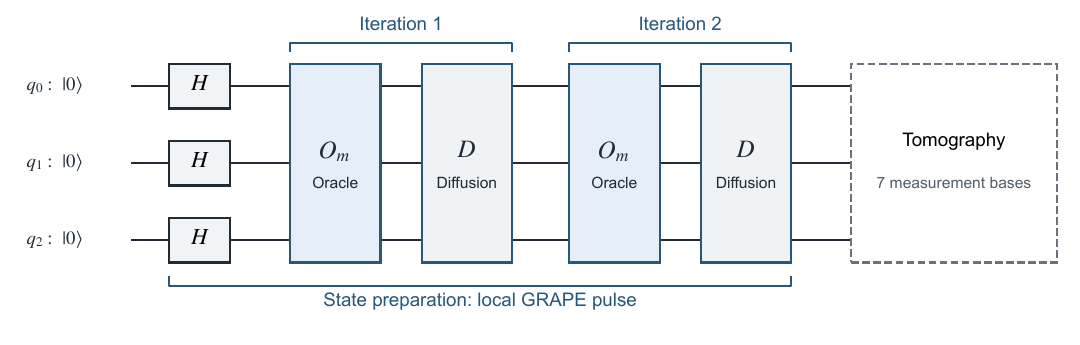}
\caption{\textbf{Three-qubit Grover task.} Two oracle--diffusion iterations amplify marked state \(m\) from a uniform superposition. The bracketed circuit defines the target of the locally optimized gradient ascent pulse engineering (GRAPE) state-preparation pulse.}
\label{fig:grover_circuit}
\end{figure}

The common logical task was three-qubit Grover state preparation with one marked state among eight computational-basis states \citep{grover1997,das2003grover}. The logical circuit in Fig.~\ref{fig:grover_circuit} starts from \(|000\rangle\) and applies a Hadamard gate to each qubit to prepare
\(
|s\rangle=H^{\otimes3}|000\rangle=8^{-1/2}\sum_{x\in\{0,1\}^3}|x\rangle.
\)
For a marked state \(|m\rangle\), one Grover iteration was \(G_m=D O_m\), where \(O_m=I-2|m\rangle\!\langle m|\) is the phase oracle and \(D=2|s\rangle\!\langle s|-I\) is the diffusion operator. The oracle reverses the phase of the marked basis state, and the diffusion operator amplifies its amplitude. Two successive iterations define the target state
\begin{equation}
 |\psi_m^{\mathrm{ideal}}\rangle
 =G_m^2|s\rangle
 =\frac{1}{8\sqrt{2}}\left(11|m\rangle-\sum_{x\neq m}|x\rangle\right),
 \label{eq:grover_output}
\end{equation}
up to a global phase. Thus, the ideal marked-state population was \(121/128\simeq0.945\), and each of the seven non-marked states had population \(1/128\). The marked states were \(|001\rangle\) on Q0, \(|100\rangle\) on Q1, \(|010\rangle\) on Q2, \(|111\rangle\) on Q4 and \(|011\rangle\) on Q3.

The logical circuit specified the target for device-specific GRAPE state preparation. Experimental outputs were reconstructed by tomography using seven measurement bases, and the marked-state population was extracted using the estimators described below.

\subsection*{Hamiltonian calibration and pulse compilation}

Simulation and pulse optimization used the local three-spin Hamiltonian
in the frame rotating at the RF carrier frequency,
\begin{equation}
 \frac{H_{\mathrm{int}}}{\hbar}
 =2\pi\left[
 \sum_{i=0}^{2}\Delta\nu_i I_z^{(i)}
 +\sum_{i<j}J_{ij}\left(
 I_x^{(i)}I_x^{(j)}+I_y^{(i)}I_y^{(j)}+I_z^{(i)}I_z^{(j)}
 \right)\right],
 \label{eq:hamiltonian}
\end{equation}
using the same convention as Supplementary Eq.~\eqref{eq:nmr_internal_hamiltonian}.
The indices \(i,j\in\{0,1,2\}\) label the three spins, and
\(I_\alpha^{(i)}=\sigma_\alpha^{(i)}/2\) for \(\alpha\in\{x,y,z\}\),
with identity operators on the other spins understood.
The parameters \(\Delta\nu_i\) are resonance-frequency offsets relative to the
RF carrier, and \(J_{ij}\) are signed scalar coupling constants; both are in Hz.
These parameters were estimated from local spectra acquired with
\(x\)- and \(y\)-phase excitation using a first-order weak-coupling
line-centre model (Supplementary Sec.~\ref{supp:control} and
Supplementary Table~\ref{tab:supp_hamiltonians}).

The control Hamiltonian was
\(H_{\mathrm{c}}(t)/\hbar=2\pi[b_x(t)\sum_i I_x^{(i)}
+b_y(t)\sum_i I_y^{(i)}]\),
with RF quadratures \(b_x,b_y\) in Hz.
GRAPE optimized an 11-ms pulse comprising 440 piecewise-constant
segments against the Grover target unitary
\(U_m=(DO_m)^2H^{\otimes3}\), expressed in the local model basis
\citep{khaneja2005grape}.
The unitary-overlap objective used detunings
\((-40,-20,0,20,40)\)~Hz and RF-amplitude scales
\((0.95,1.00,1.05)\); the same ensemble was used for the
simulations in Fig.~\ref{fig:cross}c.
The objective function, optimizer settings and waveform export
are detailed in Supplementary Sec.~\ref{supp:control}.

\subsection*{Reconstruction and validation}

Tomography using seven measurement bases provided observations of all 63 non-identity Pauli components. For Q0, populations were corrected by constrained inversion of the empirical basis-response matrix, enforcing non-negativity and unit sum. For Figs.~\ref{fig:cross} and~\ref{fig:replication}, density matrices were reconstructed from all observations by quality-weighted least squares, subject to positive semidefiniteness and unit trace. The displayed populations were then obtained by applying the same numerical virtual purification to each reconstructed matrix [Supplementary Eq.~\eqref{eq:supp_virtual_purification}].
The measurement bases, reconstruction procedures and post-processing used to obtain the reported populations are described in Supplementary Sec.~\ref{supp:reconstruction}.

The total-variation distance between the reported and ideal population distributions was
\begin{equation}
 D_{\mathrm{TV}}(p,p^{\mathrm{ideal}})
 =\frac12\sum_x|p_x-p_x^{\mathrm{ideal}}|,
 \label{eq:tvd}
\end{equation}
where \(x\) runs over the eight computational-basis states and the ideal distribution is defined by Eq.~\eqref{eq:grover_output}.

For the later sessions in Fig.~\ref{fig:replication}, each completed session comprised three consecutive complete tomography measurements. Repeatability was judged qualitatively from the consistency of the resulting marked-state populations.
Sessions with closely agreeing results were considered satisfactory. All 15 complete measurements were included, with each session summarized by its mean and observed range.
Filled and open markers indicate satisfactory and unsatisfactory repeatability, respectively.
Session records are provided in Supplementary Sec.~\ref{supp:validation}.

\bibliographystyle{naturemag}
\bibliography{references}

\section*{Author contributions}
K.L., S.-Y.H. and J.X. conceived the study. K.L. formulated the experimental strategy, and S.-Y.H. performed the experiments. Z.W., C.Z., G.F. and J.X. provided the experimental hardware and technical support. All authors contributed to writing and revising the manuscript.

\section*{Competing interests}
The authors declare no competing interests.

\input{supplementary_appendix}

\end{document}

%% file: supplementary_appendix.tex
% Generated from versioned records by figure_scripts/build_supplementary_appendix.py.
\clearpage
\appendix
\setcounter{section}{0}
\setcounter{equation}{0}
\setcounter{figure}{0}
\setcounter{table}{0}
\setcounter{page}{1}
\renewcommand{\thesection}{S\arabic{section}}
\renewcommand{\thesubsection}{\thesection.\arabic{subsection}}
\renewcommand{\theequation}{S\arabic{equation}}
\renewcommand{\thefigure}{S\arabic{figure}}
\renewcommand{\thetable}{S\arabic{table}}
\renewcommand{\thepage}{S\arabic{page}}
\renewcommand{\theHsection}{supp.\arabic{section}}
\renewcommand{\theHequation}{supp.\arabic{equation}}
\renewcommand{\theHfigure}{supp.\arabic{figure}}
\renewcommand{\theHtable}{supp.\arabic{table}}
\setstretch{1.15}
\begin{center}
{\Large\bfseries Supplementary Information}\par\medskip
Agent-assisted experimental strategy transfer across quantum processors
\end{center}

\section{Experimental platform and initial-state preparation}
\label{supp:platform}

Experiments were performed on five three-qubit liquid-state nuclear magnetic
resonance (NMR) quantum processors, \texttt{Triangulumpro}, from SpinQ Technology Co., Ltd. 
The five processors are distinguished by their hardware labels in Table~\ref{tab:supp_devices}; lowercase \(q_0,q_1,q_2\) denote spins within a processor.

\begin{table}[!htbp]
\centering\small
\caption{Processor identifiers and task assignments.}
\label{tab:supp_devices}
\setlength{\tabcolsep}{4pt}
\begin{tabularx}{\linewidth}{l l l >{\raggedright\arraybackslash}X}
\toprule
Processor & Serial & Marked state & Role \\
\midrule
Q0 & Not specified & \(|001\rangle\) & Strategy development \\
Q1 & 120058 & \(|100\rangle\) & First deployment \\
Q2 & 120054 & \(|010\rangle\) & Second deployment \\
Q4 & 120051 & \(|111\rangle\) & Third deployment \\
Q3 & 120060 & \(|011\rangle\) & Fourth deployment \\
\bottomrule
\end{tabularx}
\end{table}

\paragraph*{Instrument and sample identity.}

The built-in molecular system is iodotrifluoroethylene,
\(\mathrm{C_2F_3I}\) (\(\mathrm{F_2C{=}CFI}\)).  Its three inequivalent
spin-\(1/2\) \(^{19}\mathrm{F}\) nuclei encode the computational spins.  A
separate channel labelled H acquired a \(^{1}\mathrm{H}\) frequency
reference for field tracking and did not encode a computational qubit. 

\begin{figure}[!htbp]
\centering
\includegraphics[width=\linewidth]{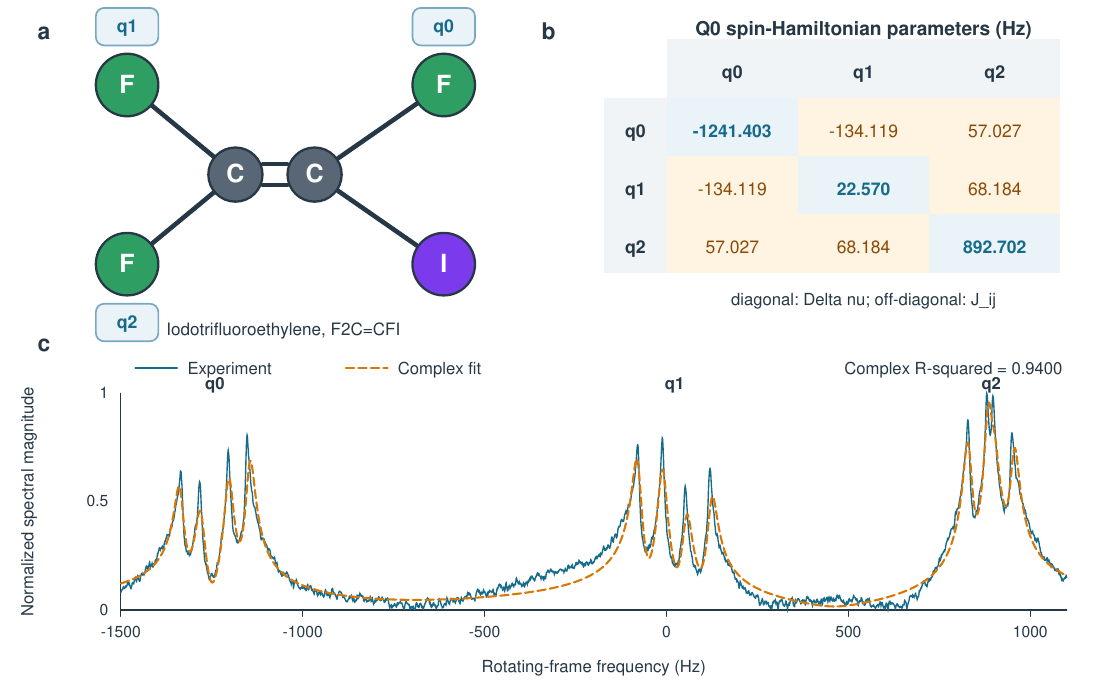}
\caption{\textbf{Molecular spin system and Q0 Hamiltonian characterization.}
\textbf{a}, Molecular structure of iodotrifluoroethylene.  In the zero-based
ordering used throughout this work, \(q_0\) is the fluorine bound to the CFI
carbon and \(q_1,q_2\) are the two fluorines bound to the CF\(_2\) carbon.
\textbf{b}, Q0 rotating-frame resonance offsets (bold diagonal, Hz) and signed
scalar couplings (off-diagonal, Hz) used for pulse design.  \textbf{c},
Measured thermal-equilibrium \(^{19}\mathrm{F}\) spectrum following a hard-X90
readout on Q0 (22 July 2026, 15:28:53), with no PPS initialization.  Blue shows
the magnitude of the recorded complex Fourier spectrum and orange the magnitude
of a complex least-squares lineshape fit.  Twelve Lorentzian transition centres
were fixed from the parameters in b; independent complex amplitudes, one linewidth
per spin and a complex linear baseline were fitted over \([-1500,1100]\)~Hz.
Both curves use the same normalization, with no smoothing of the observations.
The complex-data \(R^2\) was 0.9400.  Fitted effective FWHM values were
29.0, 29.6 and 27.9~Hz for \(q_0,q_1,q_2\), respectively; these are spectral
widths rather than independent measurements of homogeneous \(T_2\).}
\label{fig:supp_spin_system}
\end{figure}

Figure~\ref{fig:supp_spin_system} summarizes the molecular assignment, the Q0
pulse-design Hamiltonian, and a thermal-spectrum fit with transition centres fixed
by these parameters, in the
same visual organization commonly used for NMR quantum processors
\citep{das2015wsuperposition,feng2022triangulum}.  The diagonal entries are the
rotating-frame resonance offsets \(\Delta\nu_i\), conventionally called
chemical-shift frequencies in NMR quantum-computing Hamiltonians, whereas the
off-diagonal entries are the signed scalar couplings \(J_{ij}\).  In the frame rotating at the RF carrier frequency,
the internal Hamiltonian used for three-spin simulation and pulse optimization is
\begin{equation}
\frac{H_{\mathrm{int}}}{\hbar}
=
2\pi\left[
\sum_{i=0}^{2}\Delta\nu_i I_z^{(i)}
+
\sum_{i<j}J_{ij}\left(
I_x^{(i)}I_x^{(j)}+I_y^{(i)}I_y^{(j)}+I_z^{(i)}I_z^{(j)}
\right)
\right],
\label{eq:nmr_internal_hamiltonian}
\end{equation}
where \(I_\alpha^{(i)}=\sigma_\alpha^{(i)}/2\) for \(\alpha\in\{x,y,z\}\), with identity operators
on the other spins understood. The quantities \(\Delta\nu_i\)
are resonance-frequency offsets relative to the rotating-frame
carrier, and \(J_{ij}\) are scalar coupling constants; both
are expressed in Hz.

The SpinQ TriangulumPro instruments used here lack sample and magnet temperature
control, so the resonance frequencies may vary over time.

\paragraph*{Physical and effective initial states.}
Before each fluorine experiment, a nominal \(15~\mathrm{s}\) recycle interval
was used.  Under the high-temperature approximation, \(k_{\mathrm B}T\gg
h f_{\mathrm F}\), the physical thermal-equilibrium state is well approximated by
\begin{equation}
 \rho_{\mathrm{th}}^{\mathrm{phys}}
 \simeq \frac{\mathbb{I}_8}{8}
 + \epsilon_{\mathrm{th}}\,\Delta\rho_{\mathrm{th}},
 \qquad
 \Delta\rho_{\mathrm{th}} \propto
 I_z^{(0)}+I_z^{(1)}+I_z^{(2)},
 \qquad \epsilon_{\mathrm{th}}\ll 1.
 \label{eq:thermal-deviation-state}
\end{equation}
The identity term is NMR-silent and is not removed physically.  The initial state
used for quantum algorithms was instead a pseudo-pure state (PPS).  Following the
published Triangulum protocol, the platform prepares the PPS by a relaxation
method that alternates a basis-permutation operation with a \(T_1\)-relaxation
delay \citep{feng2022triangulum}.  The permutation is implemented as a GRAPE
pulse: it leaves \(|000\rangle\) unchanged while cyclically permuting the
populations of the other seven computational basis states.  Repetition of the
permutation--relaxation cycle drives those seven populations towards a common,
smaller value while \(|000\rangle\) remains dominant.  The target deviation
operator is
\begin{equation}
 \Delta\rho_{\mathrm{PPS}}
 \propto |000\rangle\!\langle 000|-\frac{\mathbb{I}_8}{8}.
 \label{eq:pps-deviation-state}
\end{equation}
Thus PPS preparation does not cool the physical ensemble into a pure state.  It
produces a deviation density operator that evolves and is detected as if the
effective state were \(|000\rangle\).  Subsequent computational-basis initial
states are generated from this PPS by applying a selective \(Y(\pi)\) pulse to
each qubit whose desired logical value is one.

During each \(15~\mathrm{s}\) fluorine recycle interval,
\(^{1}\mathrm{H}\)-reference acquisitions were serialized on the same instrument
connection at nominal \(3~\mathrm{s}\) spacing.  The most recent accepted
\(^{1}\mathrm{H}\) observation was used to infer the fluorine carrier before the
next F pulse.  The H and F operations therefore did not overlap on the shared
transmit/acquisition path.

\paragraph*{FID processing and normalization.}
Each F acquisition was recorded as a complex free-induction decay (FID),
apodized, Fourier transformed, phase corrected, and fitted to the twelve allowed
single-quantum transitions (four lines for each fluorine spin).  Three same-run
PPS/Y90 reference spectra supplied one complex PPS-reference scale \(g_q\) for
each spin.  In compact notation, the scale was obtained by complex least squares,
\begin{equation}
 g_q =
 \frac{\mathbf{s}_{q,\mathrm{pred}}^{\dagger}
       \mathbf{s}_{q,\mathrm{ref}}}
      {\mathbf{s}_{q,\mathrm{pred}}^{\dagger}
       \mathbf{s}_{q,\mathrm{pred}}},
 \qquad
 \widetilde{\mathbf{s}}_q=
 \frac{\mathbf{s}_{q,\mathrm{meas}}}{g_q},
 \label{eq:complex-pps-normalization}
\end{equation}
which simultaneously normalizes reference amplitude and receiver phase.  The
normalized real and imaginary line amplitudes from seven readout settings were
then jointly fitted in the complete three-spin Pauli basis.  The current analysis uses a quality-weighted
least-squares estimate for the 63 non-identity Pauli coefficients, adds the fixed
identity coefficient \(1/8\), and performs a trace-one positive-semidefinite
constrained fit.  The resulting constrained-fit effective-state populations are
\begin{equation}
 p_i^{\mathrm{eff}}
 =\langle i|\rho_{\mathrm{eff}}|i\rangle,
 \qquad i\in\{000,001,\ldots,111\},
 \qquad \sum_i p_i^{\mathrm{eff}}=1.
 \label{eq:effective-populations}
\end{equation}
They are normalized PPS-referenced populations, not the absolute thermodynamic
populations of individual molecules.  Under the usual effective-state
decomposition, an unmeasured deviation scale \(\eta_{\mathrm{PPS}}\) would relate
the effective estimate to the physical ensemble as
\begin{equation}
 p_i^{\mathrm{phys}}
 =\frac{1}{8}+\eta_{\mathrm{PPS}}
  \left(p_i^{\mathrm{eff}}-\frac{1}{8}\right).
 \label{eq:physical-effective-populations}
\end{equation}
Because \(\eta_{\mathrm{PPS}}\) was neither measured nor retained in the run packages, no
claim of absolute molecular-state population is made.

For comparison, the legacy SpinQuasarServer computes each Pauli coefficient from
a signed sum of four fitted transition amplitudes divided by
\(8M_q\exp(-t/T_{2,q})\), fixes the identity coefficient to \(1/8\), constructs
the density matrix, and projects it to the positive-semidefinite set.  Its
quantity labelled fidelity is the unsquared Uhlmann/root fidelity; for a pure
target this equals \(\sqrt{p_{\mathrm{target}}^{\mathrm{eff}}}\), not the target
population itself.  The constrained-fit populations in
Eq.~\eqref{eq:effective-populations}, the dominant-eigenstate-projected
populations plotted in the main figures, the historical PSD validation values
and the Q0 response-corrected values are therefore distinguished explicitly in
Supplementary Sec.~\ref{supp:reconstruction}.  ``SpinQ-compatible'',
bracketed-\(|000\rangle\), and perfect-PPS-corrected values are retained only as
secondary reconstruction or sensitivity analyses and must be labelled as such.

\section{Agent configuration and transfer decisions}
\label{supp:agent}

\subsection{Agent implementation and decision policy}
The Q0 records identify the agent as Codex based on GPT-5, although the exact model build was unavailable. The recorded software environment comprised Python 3.11.15, NumPy 1.26.4 and SciPy 1.11.4 on Windows, with a locally bundled \texttt{spinqlink} interface. These versions describe the Q0 environment; they were not independently verified for every subsequent session. Model sampling settings and a complete conversation prompt were not preserved.

The recorded policy specifies the decision cycle, transfer constraints, quality criteria and stopping rules, with subsequent changes documented in device-specific revision records. At each step, the agent assessed observations and data quality, identified missing information, selected an allowed action and evaluated the resulting evidence before continuing, requesting intervention or stopping. The policy required records of the observations, candidate actions, selected action and rationale, tool outputs, hardware results, updated assessment, acquisition cost and operator involvement.

\subsection{Transfer constraints and operator responsibilities}
Transferable context included the three-spin model class, generic experimental procedures, quality criteria, acquisition budgets, stopping rules and deterministic tools. Available actions included hard-pulse spectroscopy with \(x\)- and \(y\)-phase excitation, selective reference measurements, dedicated relaxation measurements, tomography, gradient ascent pulse engineering (GRAPE) optimization and repeated acquisition. An X/Y pair denotes two successive spectra acquired with the two excitation phases. Device-specific numerical parameters, spectra, response matrices, reference shapes and waveforms from Q0 were excluded from the prior information supplied for a new processor. These restrictions were procedural; automatic isolation of sessions or file access was not established. On Q1, a legacy fitting routine used Q0 settings in violation of this restriction, and its result was rejected.

Operators switched processors, adjusted and shimmed instruments, restored instrument servers and authorized revisions and reruns. They also provided information about instrument operating conditions. The agent assessed the spectral evidence, selected characterization procedures and coordinated quality checks, while deterministic software fitted parameters, generated waveforms and reconstructed states. Table~\ref{tab:supp_agent_cases} summarizes representative decisions and operator involvement.

\begin{table}[!htbp]
\centering\small
\caption{Recorded decisions and operator involvement.}
\label{tab:supp_agent_cases}
\setlength{\tabcolsep}{4pt}
\begin{tabularx}{\linewidth}{l >{\raggedright\arraybackslash}X >{\raggedright\arraybackslash}X}
\toprule
Case & Observation and agent response & Outcome and operator involvement \\
\midrule
Q1 & A legacy fit used Q0 settings and was rejected. Subsequent spectral drift led the agent to fit X and Y spectra separately. & Operators readjusted the instrument and authorized revisions. Fresh local calibration and reference checks were followed by successful final validation. \\
Q2 & Reference checks alternated between pass and fail. The agent required two consecutive passing readiness checks and updated the references. & After operator readjustment, local calibration and reference checks passed, supporting pulse compilation and successful final validation. \\
Q4 & Overlapping inner transitions required an adapted line-assignment method. None of nine X/Y pairs passed all initial characterization checks, so the agent stopped acquisition. & Following operator adjustment, an independent check gave a maximum frequency difference of 0.1112~Hz after removing the common shift and a maximum coupling difference of 0.0889~Hz; both were below 2~Hz. Final validation passed after pulse reoptimization and complete tomography repeats. \\
Q3 & Fresh spectra showed the same overlapping-peak pattern. The agent reused the adapted method and combined five qualified X/Y pairs. & The model, references and task pulse were established locally for Q3, and final validation passed. \\
Q3-R1/R2 & A server interruption halted Q3-R1. Acquisition later resumed, and all three complete repeats were retained. & Q3-R1 retained its historical FAIL verdict under the original reconstruction criteria. The operator authorized Q3-R2 as a separate same-day auxiliary session that did not replace Q3-R1. \\
\bottomrule
\end{tabularx}
\end{table}

\subsection{Method adaptation on Q4 and reuse on Q3}
\label{supp:adaptation_case}
\begin{figure}[!htbp]
\centering
\includegraphics[width=\linewidth]{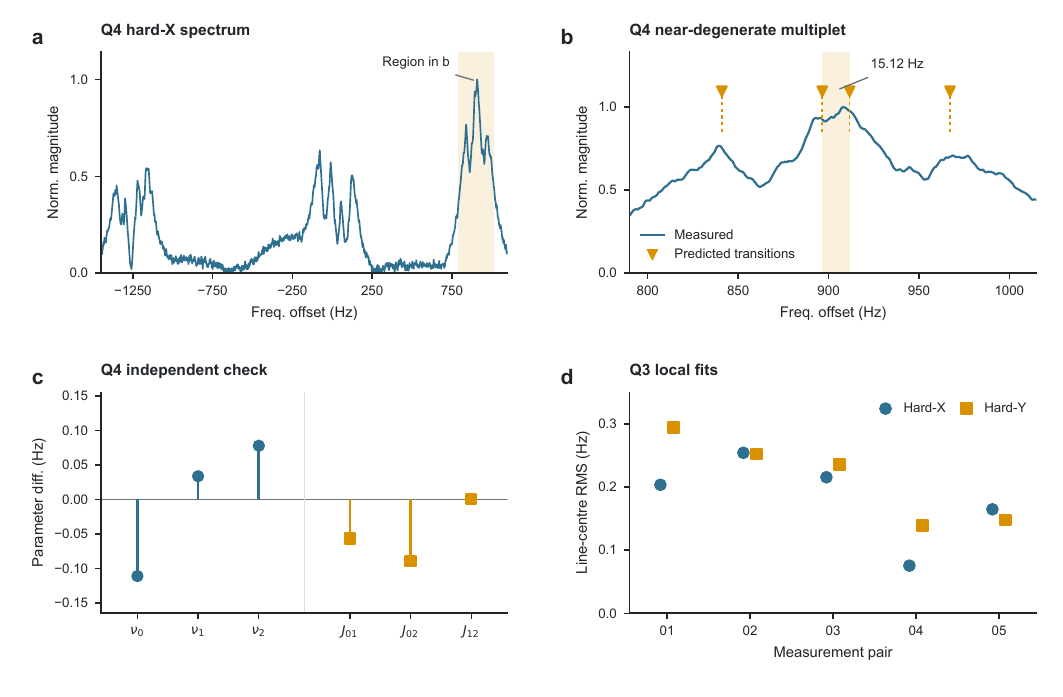}
\caption{\textbf{Local spectral evidence and model checks on Q4 and Q3.} \textbf{a}, Normalized Q4 hard-X spectral magnitude from the fourth X/Y spectral pair in the initial campaign; shading marks the region enlarged in b. \textbf{b}, The same spectrum with four predicted transition frequencies from that pair's local model; the inner separation is 15.125~Hz. \textbf{c}, Independent-check minus discovery parameters in the later Q4 rerun, with the common frequency shift removed from \(\Delta\nu_i\). The discovery model remained fixed during this check. \textbf{d}, Line-centre root-mean-square (RMS) fitting errors for the hard-X/Y spectra of five Q3 X/Y spectral pairs, all used to construct the local model.}
\label{fig:supp_adaptation}
\end{figure}

Initial Q4 X/Y spectra contained two resolved four-line multiplets and a third multiplet with overlapping inner transitions (Supplementary Fig.~\ref{fig:supp_adaptation}a,b). The original fitter required three resolved quartets. The agent therefore adapted the line-assignment procedure to infer couplings from the two resolved quartets and use the outer lines of the third multiplet as additional constraints. X and Y spectra were fitted separately to assess the fit of each spectrum and parameter consistency between acquisitions.

The adapted method was applied during the initial Q4 campaign, but none of its nine X/Y pairs passed all discovery checks. For pair 04, the X/Y line-centre root-mean-square (RMS) errors were 1.204 and 1.008~Hz, below the 2-Hz limit. However, the maximum frequency change after removing the common shift and the maximum coupling difference were 2.653 and 2.614~Hz, respectively, exceeding their 2-Hz limits. The agent stopped at the 18-acquisition budget and requested instrument adjustment. Following operator adjustment, a fresh rerun established a local model and passed an independent model check before pulse compilation and validation (Supplementary Fig.~\ref{fig:supp_adaptation}c).

On Q3, the agent applied the adapted procedure to fresh spectra with the same overlapping-peak pattern. Five X/Y pairs passed the line-centre, coupling-consistency, cross-phase and multiplet-structure checks; the median of each fitted parameter across the five pairs supplied the Q3 design model (Supplementary Fig.~\ref{fig:supp_adaptation}d). Measurement references were established locally, and the Grover pulse was reoptimized for this model. The policy entering Q3, denoted \(P_3\), was reconstructed from the Q4/Q3 records because a complete record of the agent's context at Q3 entry was unavailable.

\section{Hamiltonian estimation and pulse optimization}
\label{supp:control}

\subsection{Parameter estimation}
Table~\ref{tab:supp_hamiltonians} lists the device-local design models. The first-order line-centre model used for resolved multiplets was
\begin{equation}
 f_j(\mathbf z)=\Delta\nu_j+\frac12\sum_{k\ne j}J_{jk}z_k,\qquad z_k\in\{-1,+1\}.
 \label{eq:supp_lines}
\end{equation}
Here \(f_j\) is relative to the RF carrier and \(z_k\) is an eigenvalue of \(2I_z^{(k)}\). Spectral fitting used this weak-coupling approximation, whereas propagation retained the full scalar coupling in Eq.~\eqref{eq:nmr_internal_hamiltonian}. Q0 used 12 assigned lines with \(J_{01}<0\), \(J_{02}>0\) and \(J_{12}>0\). Later fits treated X and Y spectra separately; Q2 and Q3 used parameter-wise medians over three and five qualified pairs, respectively. The Q4/Q3 treatment of overlapping transitions is described in Supplementary Sec.~\ref{supp:adaptation_case}. Reference-only recentering did not imply pulse reoptimization with refitted couplings.

Supplementary Fig.~\ref{fig:supp_adaptation} shows Q4 spectra and model checks, and Q3 fitting diagnostics. In a,b, spectral magnitude was normalized as \(\max\{[M(f)-B]/[M_{\max}-B],0\}\), with \(B\) the second percentile and \(M_{\max}\) the maximum over the full spectrum. In c, frequency differences are \(d_j-\bar d\), where \(d_j=\Delta\nu_j^{\mathrm{held}}-\Delta\nu_j^{\mathrm{disc}}\) and \(\bar d=-1.3013\)~Hz; couplings are compared directly. The independent check passed its internal quality criteria and the limits of 20~Hz for the common shift and 2~Hz for the remaining frequency and coupling differences (maxima 0.1112 and 0.0889~Hz). Q3 line-centre RMS errors ranged from 0.075 to 0.294~Hz.

\begin{table}[!htbp]
\centering\small
\caption{Local parameters used in the saved pulse-design models (Hz).}
\label{tab:supp_hamiltonians}
\setlength{\tabcolsep}{4pt}
\begin{tabular*}{\linewidth}{@{\extracolsep{\fill}}l r r r r r r}
\toprule
Processor & \(\Delta\nu_0\) & \(\Delta\nu_1\) & \(\Delta\nu_2\) & \(J_{01}\) & \(J_{02}\) & \(J_{12}\) \\
\midrule
Q0 & -1241.403 & 22.570 & 892.702 & -134.119 & 57.027 & 68.184 \\
Q1 & -1249.818 & 37.055 & 922.326 & -130.854 & 52.298 & 68.079 \\
Q2 & -1263.144 & 23.312 & 909.158 & -130.712 & 52.622 & 67.925 \\
Q4 & -1231.744 & 42.072 & 918.078 & -130.089 & 52.358 & 67.711 \\
Q3 & -1256.425 & 31.035 & 916.586 & -129.832 & 51.568 & 67.410 \\
\bottomrule
\end{tabular*}
\par\smallskip\begin{minipage}{\linewidth}\footnotesize Rounded pulse-design parameters; full precision is retained in the source data. Parameters could vary between sessions.\end{minipage}
\end{table}

\subsection{Control objective and implementation}
The compiler expressed \(U_m=(DO_m)^2H^{\otimes3}\) in the local model basis as \(U_{\mathrm{target}}\). For common detuning \(\delta\) (Hz) and RF-amplitude scale \(s\), propagation used
\begin{equation}
 U_{\delta,s}=\prod_{\ell=M}^{1}\exp\!\left[-i\Delta t\left(\frac{H_{\mathrm{int}}}{\hbar}+2\pi\delta\sum_jI_z^{(j)}+s\frac{H_{{\rm c},\ell}}{\hbar}\right)\right],
 \label{eq:supp_propagation}
\end{equation}
with \(M=440\) and \(\Delta t=25\,\mu\mathrm{s}\). GRAPE maximized
\begin{equation}
 F_U=\sum_{\delta,s}w_{\delta,s}\frac{|\operatorname{Tr}(U_{\delta,s}U_{\mathrm{target}}^\dagger)|}{8},
 \qquad \sum_{\delta,s}w_{\delta,s}=1.
 \label{eq:supp_objective}
\end{equation}
This unitary-overlap score is distinct from measured gate fidelity. Each RF quadrature was bounded by \(0.4b_{\mathrm{cal}}\simeq3{,}333.33\)~Hz, with \(b_{\mathrm{cal}}=8{,}333.33\)~Hz and both endpoint segments fixed to zero. Optimization used SciPy's \texttt{trust-constr} with analytic gradients, at most 20 inner iterations per outer iteration, and \texttt{xtol}=\texttt{gtol}=\(10^{-4}\). Processor-specific settings are listed in Table~\ref{tab:supp_grape}.

\begin{table}[!htbp]
\centering\small
\caption{Saved settings for the deployment pulse optimizations.}
\label{tab:supp_grape}
\setlength{\tabcolsep}{4pt}
\begin{tabularx}{\linewidth}{l r >{\raggedright\arraybackslash}X r >{\raggedright\arraybackslash}X}
\toprule
Processor & Seed & Initialization & \shortstack{Maximum outer\\iterations} & Stopping criteria \\
\midrule
Q1 & 202607230 & Local checkpoint & 4 & \(F_U\geq0.97\); patience 2 \\
Q2 & 202607242 & Cold start & 12 & \(F_U\geq0.97\); patience 3 \\
Q4 & 202607251 & Cold start & 12 & Protocol: \(F_U\geq0.995\) \\
Q3 & 202607253 & Cold start & 12 & \(F_U\geq0.97\); patience 3 \\
\bottomrule
\end{tabularx}
\par\smallskip\begin{minipage}{\linewidth}\footnotesize Q1 resumed its own checkpoint. Optimization stopped at the score threshold, outer-iteration limit or configured stagnation limit. For Q1--Q3, patience counts consecutive outer iterations with improvement below \(10^{-5}\). Q4 settings come from its pre-optimization protocol; a complete final summary was unavailable. Score thresholds are configured goals, not measured populations or guarantees of convergence. All devices used the five-detuning, three-RF-scale ensemble in Methods.\end{minipage}
\end{table}

Controls were exported as quantized amplitude--phase segments, with \(b_x=b\cos\phi\) and \(b_y=b\sin\phi\). Q1 simulations using the exported waveform reproduced the archived minimum, mean and zero-detuning population at unit RF scale. The simulation grid also served as the optimization ensemble, so these results quantify sensitivity within that ensemble rather than independent out-of-ensemble robustness.

\section{Tomography and reconstruction sensitivity}
\label{supp:reconstruction}

\subsection{Measurement settings and normalization}
The seven analysis settings in Table~\ref{tab:supp_settings} each supplied 24 observations, giving 168 observations covering all 63 non-identity three-qubit Pauli components, with repeated measurements of some components. After applying the spectral sign conventions and qubit mapping, the normalized expectations define the effective state through
\begin{equation}
 \rho_{\mathrm{eff}}=\frac18\left[\mathbb I_8+\sum_{P\ne\mathbb I_8}\langle P\rangle_{\mathrm{eff}}P\right],
 \label{eq:supp_pauli_expansion}
\end{equation}
where \(P\in\{I,X,Y,Z\}^{\otimes3}\). The signal normalization is described in Supplementary Sec.~\ref{supp:platform}.

\begin{table}[!htbp]
\centering\small
\caption{Seven-setting analysis-pulse sequence.}
\label{tab:supp_settings}
\setlength{\tabcolsep}{4pt}
\begin{tabularx}{\linewidth}{*{4}{>{\raggedright\arraybackslash}X}}
\toprule
Order & \(q_0\) & \(q_1\) & \(q_2\) \\
\midrule
1 & I & I & I \\
2 & X90 & X90 & X90 \\
3 & I & I & Y90 \\
4 & X90 & Y90 & X90 \\
5 & Y90 & I & I \\
6 & X90 & X90 & Y90 \\
7 & I & Y90 & Y90 \\
\bottomrule
\end{tabularx}
\par\smallskip\begin{minipage}{\linewidth}\footnotesize X90 and Y90 denote nominal selective \(90^\circ\) rotations about the corresponding axes; I denotes no analysis pulse. Columns follow the qubit order used in reconstruction.\end{minipage}
\end{table}

\subsection{Population correction and constrained reconstruction}
For Q0, the measured population vectors for the eight nominal basis inputs form the columns of the empirical response matrix \(A\). A reconstructed population vector \(\mathbf p\) was corrected by
\begin{equation}
 \mathbf p_{\mathrm{corr}}=\underset{\mathbf z\geq0,\,\mathbf1^T\mathbf z=1}{\operatorname{argmin}}
 \|A\mathbf z-\mathbf p\|_2^2.
 \label{eq:supp_response_inverse}
\end{equation}
This population correction includes preparation, reconstruction and readout effects; it does not correct off-diagonal matrix elements.

The historical estimator selected repeated observations by proximity to the ideal-target expectations, then clipped negative eigenvalues and normalized the trace. Its results are retained to document the original terminal-validation decisions. The target-independent reconstruction used for the current figures instead fits all observations:
\begin{equation}
 \underset{\rho_{\mathrm{eff}}\succeq0,\,\operatorname{Tr}\rho_{\mathrm{eff}}=1}{\operatorname{minimize}}
 \sum_a w_a\big[\operatorname{Tr}(P_a\rho_{\mathrm{eff}})-y_a\big]^2,
 \qquad w_a=\max(R_a,0.05)^2.
 \label{eq:supp_constrained_fit}
\end{equation}
Here \(a\) indexes observations, \(P_a\) is the observed Pauli operator, \(y_a\) its normalized expectation and \(R_a\) the spectral fit score. The implementation uses accelerated projected-gradient iterations with eigenvalues projected onto the probability simplex, enforcing positive semidefiniteness (PSD) and unit trace. It stops when the successive-iterate matrix norm falls below \(10^{-11}\) or after 4,000 iterations. The target label is used only to evaluate the reconstructed result.

\subsection{Uniform numerical virtual purification}
Several reconstructed effective states showed appreciable mixedness. Tomography provided full effective density-matrix estimates, enabling purification to be performed numerically. For Figs.~\ref{fig:cross} and~\ref{fig:replication}, we applied the same numerical virtual purification to every effective density matrix reconstructed from the original measurements. Here, we use the dominant-eigenstate limit of virtual distillation~\citep{huggins2021virtual}, projecting onto the normalized eigenvector \(|v_{\max}\rangle\) with the largest eigenvalue:
\begin{equation}
 \rho_{\mathrm{vp}}=|v_{\max}\rangle\langle v_{\max}|,
 \qquad p_x^{\mathrm{vp}}=\langle x|\rho_{\mathrm{vp}}|x\rangle
 =|\langle x|v_{\max}\rangle|^2.
 \label{eq:supp_virtual_purification}
\end{equation}
This numerical post-processing yields unit purity by construction; it does not alter the measured FIDs or physically purify the ensemble. All four terminal matrices and all 15 complete later repeats were included.

Tables~\ref{tab:supp_terminal_estimators} and~\ref{tab:supp_repeats} compare the marked-state populations before and after this step. The marked state had the largest population in every matrix both before and after purification. Agreement with the ideal population and the spread across repeats nevertheless depend on this post-processing.

\begin{table}[!htbp]
\centering\small
\caption{Terminal marked-state populations before and after numerical virtual purification.}
\label{tab:supp_terminal_estimators}
\setlength{\tabcolsep}{4pt}
\begin{tabular*}{\linewidth}{@{\extracolsep{\fill}}l r r}
\toprule
Processor & Constrained & Purified \\
\midrule
Q1 & 0.860 & 0.929 \\
Q2 & 0.959 & 0.959 \\
Q4 & 0.826 & 0.936 \\
Q3 & 0.910 & 0.910 \\
\bottomrule
\end{tabular*}
\par\smallskip\begin{minipage}{\linewidth}\footnotesize Both columns use the same measurement data. Purified populations are those displayed in Fig.~\ref{fig:cross}.\end{minipage}
\end{table}
\begin{table}[!htbp]
\centering\small
\caption{Marked-state populations before and after numerical virtual purification for all complete later repeats.}
\label{tab:supp_repeats}
\setlength{\tabcolsep}{4pt}
\begin{tabular*}{\linewidth}{@{\extracolsep{\fill}}l r r r}
\toprule
Run & Repeat & Constrained & Purified \\
\midrule
Q2-R1 & 1 & 0.963 & 0.963 \\
Q2-R1 & 2 & 0.970 & 0.970 \\
Q2-R1 & 3 & 0.960 & 0.960 \\
Q4-R2 & 1 & 0.769 & 0.938 \\
Q4-R2 & 2 & 0.591 & 0.894 \\
Q4-R2 & 3 & 0.893 & 0.940 \\
Q4-R3 & 1 & 0.748 & 0.947 \\
Q4-R3 & 2 & 0.734 & 0.938 \\
Q4-R3 & 3 & 0.808 & 0.940 \\
Q3-R1 & 1 & 0.838 & 0.922 \\
Q3-R1 & 2 & 0.845 & 0.924 \\
Q3-R1 & 3 & 0.927 & 0.934 \\
Q3-R2 & 1 & 0.945 & 0.945 \\
Q3-R2 & 2 & 0.937 & 0.937 \\
Q3-R2 & 3 & 0.945 & 0.945 \\
\bottomrule
\end{tabular*}
\par\smallskip\begin{minipage}{\linewidth}\footnotesize All three complete repeats per session are included. Purified populations are those displayed in Fig.~\ref{fig:replication}. Q1 used an earlier single-block protocol and is listed separately in Table~\ref{tab:supp_sessions}.\end{minipage}
\end{table}

\section{Validation policies and run accounting}
\label{supp:validation}

\subsection{Task performance and repeatability}
The campaign proceeded Q1--Q2--Q4--Q3, with local models and pulses fixed before each reported final measurement. Strategy adaptation is described in Supplementary Sec.~\ref{supp:agent}. Figure~\ref{fig:cross} characterizes task performance by marked-state populations and total-variation distances from the ideal distribution, using the uniformly purified effective states defined in Supplementary Sec.~\ref{supp:reconstruction}. Table~\ref{tab:supp_terminal_estimators} compares constrained and purified populations.

For Fig.~\ref{fig:replication}, repeatability is assessed qualitatively from three complete tomography repeats per session. Closely agreeing purified marked-state populations are classified as satisfactory, without a numerical cutoff. All 15 complete repeats from five sessions are included; each session is summarized by its mean and range (maximum minus minimum). Filled and open circles denote satisfactory and unsatisfactory repeatability. These retrospective assessments are distinct from the original protocol verdicts retained in the records.

\subsection{Reference checks and acquisition sequence}
Reference measurements established instrument readiness and local reconstruction references. The later protocol used the spectral-quality and temporal-consistency requirements in Table~\ref{tab:supp_terminal_gates}. These checks assess measurement quality separately from task performance and repeatability.

\begin{table}[!htbp]
\centering\small
\caption{Reference quality checks for the later repeated-measurement sessions.}
\label{tab:supp_terminal_gates}
\setlength{\tabcolsep}{4pt}
\begin{tabularx}{\linewidth}{l >{\raggedright\arraybackslash}X}
\toprule
Check & Requirement \\
\midrule
Spectral quality & \(R_{\mathrm{fit}}\geq0.98\), peak-to-noise ratio \(\geq20\), relative amplitude \(\geq0.35\), relative active area \(\geq0.20\), positive signed area, and fitted centre and linewidth not reaching their fitting bounds. \\
Temporal consistency & For each qubit, absolute fractional deviations from the three-anchor medians \(\leq0.30\) for amplitude and \(\leq0.35\) for active area; absolute centre deviation \(\leq20\) Hz. \\
\bottomrule
\end{tabularx}
\end{table}

\begin{samepage}
The fit score is \(R_{\mathrm{fit}}=1-\sum_i(y_i-\hat y_i)^2/\sum_i y_i^2\), where \(y_i\) and \(\hat y_i\) are measured and fitted spectral values. Peak-to-noise ratio is the maximum spectral magnitude divided by the median of the lower half of the spectral magnitudes. Relative amplitude and active area are normalized to their respective three-channel medians within each reference block.\par
\end{samepage}

Before target acquisition, references were collected at three time points (anchors), with three spectra per qubit per anchor. Each qubit required at least two passing spectra. Anchor metrics were medians over passing spectra; at least two anchors had to meet the consistency limits for all three qubits. Reconstruction references combined all three scheduled spectra per qubit after frequency and phase alignment. Only consistent anchors supported interpolation, with no extrapolation beyond their time window.

Three further anchors bracketed target acquisition: PRE before tomography repeat 1, MID between repeats 1 and 2, and POST after repeat 3, using the same reference rules. Each complete repeat comprised seven settings. The planned session total was therefore \(27+27+21=75\) acquisitions: initial references, PRE/MID/POST references and tomography, respectively. Q2-R1 and completed Q3-R1 records confirm these counts.

\subsection{Session completion and retained attempts}
Table~\ref{tab:supp_sessions} counts complete seven-setting repeats, independently of historical reconstruction-quality verdicts. Q1 used an earlier single-repeat protocol. Q4-R1 stopped after reference failure before target acquisition and contributes no target-state population.

\begin{table}[!htbp]
\centering\small
\caption{Session completion and qualitative repeatability.}
\label{tab:supp_sessions}
\setlength{\tabcolsep}{4pt}
\begin{tabularx}{\linewidth}{l l r r >{\raggedright\arraybackslash}X}
\toprule
Session & Day & Repeats & Range & Repeatability \\
\midrule
Q1-rep & 1 & 1 & --- & Not assessed: one repeat. \\
Q4-R1 & 3 & 0 & --- & Not assessed: reference failure. \\
Q4-R2 & 3 & 3 & 0.0458 & Unsatisfactory. \\
Q2-R1 & 3 & 3 & 0.0092 & Satisfactory. \\
Q3-R1 & 3 & 3 & 0.0119 & Unsatisfactory. \\
Q3-R2 & 3 & 3 & 0.0084 & Satisfactory. \\
Q4-R3 & 4 & 3 & 0.0090 & Satisfactory. \\
\bottomrule
\end{tabularx}
\par\smallskip\begin{minipage}{\linewidth}\footnotesize Day 1 is the Q1 session, consistent with Fig.~\ref{fig:replication}; day numbers preserve the elapsed calendar-day intervals. Ranges use the purified populations in Table~\ref{tab:supp_repeats}. No three-repeat range is reported for Q1 or Q4-R1.\end{minipage}
\end{table}

Q3-R1 completed all three repeats after resuming from a server interruption. Q3-R2 was an additional, operator-authorized session on the same day and does not replace Q3-R1. Earlier Q4 validation records retain one pre-reference abort and two complete but reference-invalid attempts before the reported fourth block. Unsuccessful attempts are recorded separately from reported endpoints.

The four processors are the experimental units; reference acquisitions and tomography repeats are technical measurements. Observed ranges are descriptive, not confidence intervals.

%% file: main_quantum_method.bbl
\begin{thebibliography}{10}
\expandafter\ifx\csname url\endcsname\relax
  \def\url#1{\texttt{#1}}\fi
\expandafter\ifx\csname urlprefix\endcsname\relax\def\urlprefix{URL }\fi
\providecommand{\bibinfo}[2]{#2}
\providecommand{\eprint}[2][]{\url{#2}}

\bibitem{pompili2022network}
\bibinfo{author}{Pompili, M.} \emph{et~al.}
\newblock \bibinfo{title}{Experimental demonstration of entanglement delivery
  using a quantum network stack}.
\newblock \emph{\bibinfo{journal}{npj Quantum Inf.}}
  \textbf{\bibinfo{volume}{8}}, \bibinfo{pages}{121} (\bibinfo{year}{2022}).

\bibitem{main2025distributed}
\bibinfo{author}{Main, D.} \emph{et~al.}
\newblock \bibinfo{title}{Distributed quantum computing across an optical
  network link}.
\newblock \emph{\bibinfo{journal}{Nature}} \textbf{\bibinfo{volume}{638}},
  \bibinfo{pages}{383--388} (\bibinfo{year}{2025}).

\bibitem{li2025thread}
\bibinfo{author}{Li, K.} \emph{et~al.}
\newblock \bibinfo{title}{Realization of thread level parallelism on quantum
  devices}.
\newblock \bibinfo{howpublished}{Preprint at
  \url{https://arxiv.org/abs/2511.05436}} (\bibinfo{year}{2025}).

\bibitem{proctor2022capabilities}
\bibinfo{author}{Proctor, T.}, \bibinfo{author}{Rudinger, K.},
  \bibinfo{author}{Young, K.}, \bibinfo{author}{Nielsen, E.} \&
  \bibinfo{author}{Blume-Kohout, R.}
\newblock \bibinfo{title}{Measuring the capabilities of quantum computers}.
\newblock \emph{\bibinfo{journal}{Nat. Phys.}} \textbf{\bibinfo{volume}{18}},
  \bibinfo{pages}{75--79} (\bibinfo{year}{2022}).

\bibitem{zhu2022crossplatform}
\bibinfo{author}{Zhu, D.} \emph{et~al.}
\newblock \bibinfo{title}{Cross-platform comparison of arbitrary quantum
  states}.
\newblock \emph{\bibinfo{journal}{Nat. Commun.}} \textbf{\bibinfo{volume}{13}},
  \bibinfo{pages}{6620} (\bibinfo{year}{2022}).

\bibitem{helsen2023network}
\bibinfo{author}{Helsen, J.} \& \bibinfo{author}{Wehner, S.}
\newblock \bibinfo{title}{A benchmarking procedure for quantum networks}.
\newblock \emph{\bibinfo{journal}{npj Quantum Inf.}}
  \textbf{\bibinfo{volume}{9}}, \bibinfo{pages}{17} (\bibinfo{year}{2023}).

\bibitem{proctor2020drift}
\bibinfo{author}{Proctor, T.} \emph{et~al.}
\newblock \bibinfo{title}{Detecting and tracking drift in quantum information
  processors}.
\newblock \emph{\bibinfo{journal}{Nat. Commun.}} \textbf{\bibinfo{volume}{11}},
  \bibinfo{pages}{5396} (\bibinfo{year}{2020}).

\bibitem{moon2020tuning}
\bibinfo{author}{Moon, H.} \emph{et~al.}
\newblock \bibinfo{title}{Machine learning enables completely automatic tuning
  of a quantum device faster than human experts}.
\newblock \emph{\bibinfo{journal}{Nat. Commun.}} \textbf{\bibinfo{volume}{11}},
  \bibinfo{pages}{4161} (\bibinfo{year}{2020}).

\bibitem{reuer2023feedback}
\bibinfo{author}{Reuer, K.} \emph{et~al.}
\newblock \bibinfo{title}{Realizing a deep reinforcement learning agent for
  real-time quantum feedback}.
\newblock \emph{\bibinfo{journal}{Nat. Commun.}} \textbf{\bibinfo{volume}{14}},
  \bibinfo{pages}{7138} (\bibinfo{year}{2023}).

\bibitem{szymanski2023autonomous}
\bibinfo{author}{Szymanski, N.~J.} \emph{et~al.}
\newblock \bibinfo{title}{An autonomous laboratory for the accelerated
  synthesis of inorganic materials}.
\newblock \emph{\bibinfo{journal}{Nature}} \textbf{\bibinfo{volume}{624}},
  \bibinfo{pages}{86--91} (\bibinfo{year}{2023}).

\bibitem{canty2025sdl}
\bibinfo{author}{Canty, R.~B.} \emph{et~al.}
\newblock \bibinfo{title}{Science acceleration and accessibility with
  self-driving labs}.
\newblock \emph{\bibinfo{journal}{Nat. Commun.}} \textbf{\bibinfo{volume}{16}},
  \bibinfo{pages}{3856} (\bibinfo{year}{2025}).

\bibitem{boiko2023autonomous}
\bibinfo{author}{Boiko, D.~A.}, \bibinfo{author}{MacKnight, R.},
  \bibinfo{author}{Kline, B.} \& \bibinfo{author}{Gomes, G.}
\newblock \bibinfo{title}{Autonomous chemical research with large language
  models}.
\newblock \emph{\bibinfo{journal}{Nature}} \textbf{\bibinfo{volume}{624}},
  \bibinfo{pages}{570--578} (\bibinfo{year}{2023}).

\bibitem{alexeev2025aiquantum}
\bibinfo{author}{Alexeev, Y.} \emph{et~al.}
\newblock \bibinfo{title}{Artificial intelligence for quantum computing}.
\newblock \emph{\bibinfo{journal}{Nat. Commun.}} \textbf{\bibinfo{volume}{16}},
  \bibinfo{pages}{10829} (\bibinfo{year}{2025}).

\bibitem{cao2025agents}
\bibinfo{author}{Cao, S.} \emph{et~al.}
\newblock \bibinfo{title}{Automating quantum computing laboratory experiments
  with an agent-based {AI} framework}.
\newblock \emph{\bibinfo{journal}{Patterns}} \textbf{\bibinfo{volume}{6}},
  \bibinfo{pages}{101372} (\bibinfo{year}{2025}).

\bibitem{li2026llmqubit}
\bibinfo{author}{Li, S.} \emph{et~al.}
\newblock \bibinfo{title}{Large language model-assisted superconducting qubit
  experiments}.
\newblock \bibinfo{howpublished}{Preprint at
  \url{https://arxiv.org/abs/2603.08801}} (\bibinfo{year}{2026}).

\bibitem{xu2026vibe}
\bibinfo{author}{Xu, H.} \emph{et~al.}
\newblock \bibinfo{title}{{Vibe Calibration}: Autonomous bring-up of a
  112-qubit superconducting quantum processor by a skill-orchestrating language
  agent}.
\newblock \bibinfo{howpublished}{Preprint at
  \url{https://arxiv.org/abs/2606.22376}} (\bibinfo{year}{2026}).

\bibitem{wang2026artiq}
\bibinfo{author}{Wang, D.}, \bibinfo{author}{Qi, L.}, \bibinfo{author}{Xie,
  Y.}, \bibinfo{author}{Linke, N.~M.} \& \bibinfo{author}{Brown, K.~R.}
\newblock \bibinfo{title}{A hardware-safety-gated system for {LLM}-written
  native {ARTIQ} control code on a trapped-ion platform}.
\newblock \bibinfo{howpublished}{Preprint at
  \url{https://arxiv.org/abs/2606.27231}} (\bibinfo{year}{2026}).

\bibitem{isogawa2026agentic}
\bibinfo{author}{Isogawa, T.}, \bibinfo{author}{Okabe, R.},
  \bibinfo{author}{Phadetsuwannukun, N.}, \bibinfo{author}{Li, M.} \&
  \bibinfo{author}{Cappellaro, P.}
\newblock \bibinfo{title}{Agentic {AI} for scientific reasoning in autonomous
  quantum sensing experiments}.
\newblock \bibinfo{howpublished}{Preprint at
  \url{https://arxiv.org/abs/2607.25145}} (\bibinfo{year}{2026}).

\bibitem{vandersypen2005nmr}
\bibinfo{author}{Vandersypen, L. M.~K.} \& \bibinfo{author}{Chuang, I.~L.}
\newblock \bibinfo{title}{{NMR} techniques for quantum control and
  computation}.
\newblock \emph{\bibinfo{journal}{Rev. Mod. Phys.}}
  \textbf{\bibinfo{volume}{76}}, \bibinfo{pages}{1037--1069}
  (\bibinfo{year}{2005}).

\bibitem{li2021polynomial}
\bibinfo{author}{Li, K.} \emph{et~al.}
\newblock \bibinfo{title}{Optimizing a polynomial function on a quantum
  processor}.
\newblock \emph{\bibinfo{journal}{npj Quantum Inf.}}
  \textbf{\bibinfo{volume}{7}}, \bibinfo{pages}{16} (\bibinfo{year}{2021}).

\bibitem{grover1997}
\bibinfo{author}{Grover, L.~K.}
\newblock \bibinfo{title}{Quantum mechanics helps in searching for a needle in
  a haystack}.
\newblock \emph{\bibinfo{journal}{Phys. Rev. Lett.}}
  \textbf{\bibinfo{volume}{79}}, \bibinfo{pages}{325--328}
  (\bibinfo{year}{1997}).

\bibitem{das2003grover}
\bibinfo{author}{Das, R.}, \bibinfo{author}{Mahesh, T.~S.} \&
  \bibinfo{author}{Kumar, A.}
\newblock \bibinfo{title}{Experimental implementation of {Grover}'s search
  algorithm using efficient quantum state tomography}.
\newblock \emph{\bibinfo{journal}{Chem. Phys. Lett.}}
  \textbf{\bibinfo{volume}{369}}, \bibinfo{pages}{8--15}
  (\bibinfo{year}{2003}).

\bibitem{khaneja2005grape}
\bibinfo{author}{Khaneja, N.}, \bibinfo{author}{Reiss, T.},
  \bibinfo{author}{Kehlet, C.}, \bibinfo{author}{Schulte-Herbr{\"u}ggen, T.} \&
  \bibinfo{author}{Glaser, S.~J.}
\newblock \bibinfo{title}{Optimal control of coupled spin dynamics: design of
  {NMR} pulse sequences by gradient ascent algorithms}.
\newblock \emph{\bibinfo{journal}{J. Magn. Reson.}}
  \textbf{\bibinfo{volume}{172}}, \bibinfo{pages}{296--305}
  (\bibinfo{year}{2005}).

\bibitem{dupuis2024qiskit}
\bibinfo{author}{Dupuis, N.} \emph{et~al.}
\newblock \bibinfo{title}{{Qiskit Code Assistant}: Training {LLM}s for
  generating quantum computing code}.
\newblock In \emph{\bibinfo{booktitle}{2024 {IEEE} {LLM} Aided Design Workshop
  ({LAD})}}, \bibinfo{pages}{1--4} (\bibinfo{year}{2024}).

\bibitem{jern2025agentq}
\bibinfo{author}{Jern, L.}, \bibinfo{author}{Uotila, V.}, \bibinfo{author}{Yu,
  C.} \& \bibinfo{author}{Zhao, B.}
\newblock \bibinfo{title}{{Agent-Q}: Fine-tuning large language models for
  quantum circuit generation and optimization}.
\newblock In \emph{\bibinfo{booktitle}{2025 {IEEE} International Conference on
  Quantum Computing and Engineering ({QCE})}}, \bibinfo{pages}{1621--1632}
  (\bibinfo{year}{2025}).

\bibitem{arlt2026metadesign}
\bibinfo{author}{Arlt, S.} \emph{et~al.}
\newblock \bibinfo{title}{Meta-designing quantum experiments with language
  models}.
\newblock \emph{\bibinfo{journal}{Nat. Mach. Intell.}}
  \textbf{\bibinfo{volume}{8}}, \bibinfo{pages}{148--157}
  (\bibinfo{year}{2026}).

\bibitem{feng2022triangulum}
\bibinfo{author}{Feng, G.} \emph{et~al.}
\newblock \bibinfo{title}{{SpinQ Triangulum}: A commercial three-qubit desktop
  quantum computer}.
\newblock \emph{\bibinfo{journal}{IEEE Nanotechnol. Mag.}}
  \textbf{\bibinfo{volume}{16}}, \bibinfo{pages}{20--29}
  (\bibinfo{year}{2022}).

\bibitem{das2015wsuperposition}
\bibinfo{author}{Das, D.}, \bibinfo{author}{Dogra, S.}, \bibinfo{author}{Dorai,
  K.} \& \bibinfo{author}{{Arvind}}.
\newblock \bibinfo{title}{Experimental construction of a {W} superposition
  state and its equivalence to the {Greenberger-Horne-Zeilinger} state under
  local filtration}.
\newblock \emph{\bibinfo{journal}{Phys. Rev. A}} \textbf{\bibinfo{volume}{92}},
  \bibinfo{pages}{022307} (\bibinfo{year}{2015}).

\bibitem{huggins2021virtual}
\bibinfo{author}{Huggins, W.~J.} \emph{et~al.}
\newblock \bibinfo{title}{Virtual distillation for quantum error mitigation}.
\newblock \emph{\bibinfo{journal}{Phys. Rev. X}} \textbf{\bibinfo{volume}{11}},
  \bibinfo{pages}{041036} (\bibinfo{year}{2021}).

\end{thebibliography}
